\documentclass{JHEP3}

\usepackage{amssymb}
\usepackage{amsfonts}
\usepackage{amsbsy}
\usepackage{amsmath}
\usepackage{amsthm}
\usepackage{graphicx}
\usepackage{epstopdf}
\usepackage[vcentermath]{youngtab}
\usepackage{multirow}
\usepackage{latexsym}
\usepackage{array}

\def\half{\frac{1}{2}}
\def\Z{\mathbb{Z}}

\def\C{\mathbb{C}}

\def\T{\mathbb{T}}
\def\S{\mathbb{S}}

\def\R{\mathbb{R}}

\def\P{\mathbb{P}}

\def\n3a{t}

\def\tr{{\mathrm{tr}}}

\newcommand{\be}{\begin{equation}}
\newcommand{\ee}{\end{equation}}
\newcommand{\eq}[1]{(\ref{#1})}

\title{Charge Lattices and Consistency of 6D Supergravity}

\author{Nathan Seiberg$^{1}$ and Washington Taylor$^2$\\
$^1$School of Natural Sciences\\
Institute for Advanced Study \\
Einstein Drive, Princeton, NJ 08540, USA\\
$^2$Center for Theoretical Physics\\
Department of Physics\\
Massachusetts Institute of Technology\\
Cambridge, MA 02139, USA\\
\\
\\
{\tt seiberg} {\rm at} {\tt ias.edu},
{\tt wati} {\rm at} {\tt mit.edu}
}

\preprint{MIT-CTP-4216}

\abstract{We extend the known consistency conditions on the low-energy
  theory of six-dimensional ${\cal N} = 1$ supergravity.  We
  review some facts about the theory of two-form gauge fields and
  conclude that the charge lattice $\Gamma$ for such a theory has to
  be self-dual.  The Green-Schwarz anomaly cancellation conditions in
  the supergravity theory determine a sublattice of $\Gamma$.  The
  condition that this sublattice can be extended to a self-dual
  lattice $\Gamma$ leads to a strong constraint on theories that
  otherwise appear to be self-consistent.}

\begin{document}

\section{Introduction}

The space of chiral supergravity theories in six dimensions is
strongly constrained by anomaly cancellation and other consistency
conditions.  It was shown in \cite{Seiberg-n2} that ${\cal N} = (2,
0)$ supersymmetry in six dimensions uniquely determines the massless
spectrum of the theory.  A systematic analysis of ${\cal N} = (1, 0)$
6D supergravity theories and their string realizations was initiated
in \cite{universality}-\cite{0}.  The space
of low-energy 6D supergravity theories that are compatible
with known anomaly and kinetic term sign conditions is quite
constrained, but nevertheless contains a rich range of theories with
different gauge groups and matter content.  Many of these theories can
be realized in string theory, through F-theory \cite{Vafa-f, Morrison-Vafa-I,
  Morrison-Vafa-II} and other
constructions.  There are, however, also many theories that satisfy
these constraints with no known string realization.

If we find an apparently consistent supergravity theory that cannot
be obtained in string theory we face three logical possibilities:
\begin{enumerate}
  \item This low-energy theory is inconsistent because of
    a consideration that we are not yet aware of.
  \item There is a string construction leading to this supergravity
    theory that is based on techniques we are not yet aware of.
  \item There are consistent low-energy theories that cannot be
    obtained by any string construction.  Of course, it is difficult
    to prove that this is the case because to prove this we would need
    to know all possible string constructions and understand all
    macroscopic constraints.
\end{enumerate}

A productive research direction involves looking for such examples and
trying to see which of the three possibilities above is the right one.
This line of investigation can focus our thinking and point us either
to new string constructions or to new macroscopic consistency
conditions.

In \cite{KMT, tensors}, an explicit correspondence between the
discrete data describing the field content of low-energy supergravity
theories and topological information about F-theory constructions was
used to identify some  classes of apparently consistent models
without string realizations.  One condition that is satisfied by all
known 6D ${\cal N} = 1$ string theory vacua is that the lattice of
allowed dyonic string charges in the low-energy theory must be a
unimodular (self-dual) lattice.
This raises
the question of whether such self-duality is a necessary condition for
consistency of the quantum theory.

An extremely interesting line of research starting with
\cite{Witten-5} and references therein has explored the consistency
conditions on gauge theories based on higher form gauge fields.  In
particular, \cite{Hopkins-Singer}-\cite{ Mooretalks} have presented a
general and beautiful description of such theories.  This work leads
to the conclusion that the natural framework for describing chiral
fields is in a general class of self-dual cohomology theories.
However, the discussion in these papers is fairly abstract, and more
general than needed for our purposes here.  Instead of attempting to
frame the discussion here in the more general context of that work, we
will take a lowbrow approach based on
standard methods in field theory, which leads us to the conclusions we
need.  We should emphasize, however, that many of the observations
made here are simple corollaries of the more general theory presented
in these papers.

In this note we show by elementary arguments that any consistent 6D
${\cal N} = 1$ theory must have a lattice $\Gamma$ of allowed dyonic
string charges that is unimodular; {\it i.e.}\ $\Gamma = \Gamma^*$.
This result is independent of string theory; the dyonic strings are
simply the possible states charged under the two-form fields in the
theory.  We describe the self-duality of the charge lattice in Section
\ref{sec:consistency}.  The consequences of this condition for
constraining the space of 6D supergravity theories are discussed in
Section \ref{sec:SUGRA}.  Examples of theories that are ruled out from
this consideration are given in Section \ref{sec:examples}. Section
\ref{sec:conclusions} contains concluding remarks.

\section{Consistency conditions for six-dimensional two-form gauge fields}
\label{sec:consistency}

Our goal in this section is to find some necessary conditions on the
set of charges that can couple to six-dimensional two-form gauge
fields.  As stated in the introduction, this is a lowbrow version of
the discussion in \cite{Witten-5}-\cite{Mooretalks}.  We will start by
studying two warmup problems in two and four dimensions and then we
will turn to the six-dimensional theory.  We show that a 6D field theory with self-dual and anti-self-dual
two-form fields cannot be consistent after dimensional reduction to 2D
or 4D unless the charges lie in a self-dual lattice.

\subsection{Scalar fields in  two dimensions}

The theory of free compact scalar fields in two dimensions is well
known.  Following \cite{Narain, Narain-sw} we study $p$
left moving scalars $\phi^{+a}$ and $q$ right moving scalars $\phi^{-r}$ with standard two-point functions
\begin{equation}
\langle \phi^{+a}(0) \phi^{+b}(z) \rangle = -\delta^{ab} \log z \qquad ; \qquad \langle \phi^{-r}(0) \phi^{-s}(\bar z) \rangle = -\delta^{rs} \log \bar z  ~.
\label{eq:twodprop}
\end{equation}
Typical operators in the theory are of the form
\begin{equation}
{\cal O}_{v_a,\tilde v_r}=\exp\left(i v_a \phi^{+a} + i \tilde v_r \phi^{-r}\right) ~.
\label{eq:twodop}
\end{equation}
Using (\ref{eq:twodprop}) their operator product expansion is
\begin{equation}
{\cal O}_{v_a,\tilde v_r} {\cal O}_{u_a,\tilde u_r} \sim z^{v_au_a} \bar z^{\tilde v_r \tilde u_r}{\cal O}_{v_a+u_a,\tilde v_r+\tilde u_r} + \cdots
\label{eq:twodope}
\end{equation}
Therefore, completeness of the operator product expansion shows that
$(v_a,\tilde v_r)$ are vectors in a $p+q$ dimensional lattice
$\Gamma$.  Furthermore, single valuedness of (\ref{eq:twodope}) means that
\begin{equation}
v_a u_a - \tilde v_r \tilde u_r \in {\Z}~.
\label{eq:integralla}
\end{equation}
This means that the lattice $\Gamma$ has signature $(p,q)$ and locality means that it must be an integral lattice.  Note that the bilinear form on the lattice was determined by our choice of two point functions for the scalars (\ref{eq:twodprop}).

As long as we limit ourselves to correlation functions in ${\R}^2$,
equation (\ref{eq:integralla}) is the only consistency condition.
Studying the theory on more complicated surfaces leads to further
constraints.  In particular, imposing modular invariance
on the partition function when the
theory is on ${\T}^2$ leads to two additional conditions:
\begin{enumerate}
  \item Invariance under the $S$ transformation in the modular group
    forces the lattice $\Gamma$ to be self-dual.
  \item Invariance under the $T$ transformation in the modular group
    forces the lattice $\Gamma$ to be not only integral but also even.
    Without this requirement, the torus partition function is
    invariant under $T^2$ but not under $T$.
\end{enumerate}

Should we impose these two requirements?  In the context of string theory, the proper spacetime interpretation of the theory forces us to sum over the spin structures on the worldsheet.  Hence, the torus partition function must be fully modular invariant and therefore $\Gamma$ should be even and self-dual.  However, if we are interested in a two-dimensional field theory without insisting on another interpretation of the theory in the target space, there is no need to be so restrictive.

Let us consider the situation without imposing $T$ invariance.  The
operators (\ref{eq:twodop}) with even $v_a^2 - \tilde v_r^2$ have
integer spin and they are bosons, but those with odd $v_a^2 - \tilde
v_r^2$ have half integral spin and they are fermions.  The presence of
operators with half integer spin forces us to specify the spin
structure on the torus.  If we impose invariance under $T$ the
operators (\ref{eq:twodop}) include only bosons and we do not need to
specify the spin structure on the torus.  But if the theory is not $T$
invariant, additional data, in particular the spin structure,
can
be included in the definition of the theory.  With such data there is
no need to impose $T$ invariance for the theory to be consistent.

The situation with $S$ invariance is different.  With a choice of
one-cycles we view the torus as $\S^1\times \S^1$ and interpret the
first $\S^1$  as space and the second $\S^1$ as Euclidean time.  The
standard state/operator correspondence of the operators
(\ref{eq:twodop}) leads us to choose the spin structure in which the
half-integer spin operators are anti-periodic in space.  The trace
over the Hilbert space corresponds to making these operators
anti-periodic in Euclidean time.  In this case the $S$ transformation,
which exchanges these two circles, must be a symmetry of the theory.

We conclude that the second requirement above of $T$ invariance is
optional, but we must always impose $S$ invariance.  Therefore,
$\Gamma$ must be integral and self-dual but not necessarily even.

An example of a consistent theory that is $S$ invariant but not $T$ invariant is the theory of a free left moving Weyl fermion.  In its bosonized form it fits the discussion above about the free scalars with $p=1$, $q=0$.  The corresponding lattice is integral and self-dual but it is not even.

\subsection{Maxwell fields in four dimensions}

As our second warmup example we remind the reader of the theory of free Maxwell fields in four dimensions $A^a$.  The analog of (\ref{eq:twodop}) are loop operators including Wilson loops and 't Hooft loops.  The former are of the form
\begin{equation}
W_{v_a}=\exp\left(i v_a \oint_{\Sigma_1} A^a \right)~,
    \label{eq:Wilson}
\end{equation}
where $\Sigma_1$ is a closed loop.  This operator describes the effect of a probe particle with ``electric charges'' $v_a$.  Again, the vector $v_a$ belongs to a lattice $L$.  The 't Hooft operators describe the effect of probe magnetic particles with magnetic charges $u^a$.  The Dirac quantization condition
can be understood in terms of the relative locality of Wilson
and 't Hooft operators.  It implies that
\begin{equation}
v_a u^a \in {\Z}~,
\label{eq:Diracqua}
\end{equation}
which is similar to (\ref{eq:integralla}) and states that
$u^a$ belongs to the dual lattice $L^*$.

More generally, states in the 4D theory can have dyonic charges $(v_a,
u^b)$.  The Dirac quantization condition states that for any two
dyonic charges $(v, u)$ and $(v', u')$ the antisymmetric charge
product
\begin{equation}
v\cdot u' - u \cdot v' \in \Z
\label{eq:}
\end{equation}
is integral.  This describes a skew-symmetric inner product on the charge space.  If the gauge group is $U(1)^n $, the skew-symmetric inner product can be written in the canonical form
\begin{equation}
\left(\begin{array}{cc}
0 & I\\
-I & 0
\end{array} \right)
\label{eq:4D-matrix}
\end{equation}
where $I$ is the $n \times n$ identity matrix.  Note that the matrix
(\ref{eq:4D-matrix}) has unit determinant.  The restriction of Dirac
quantization alone does not uniquely determine the form of
(\ref{eq:4D-matrix}).  Other integral skew-symmetric matrices might be
considered for describing charges of candidate theories.  Such models
are incompatible, however, with the condition that Wilson loops and 't
Hooft loops are the proper observables for the theory.  {}From
representation theory of the relativistic symmetry group, in 4D these
are the only available classes of operators.  We discuss below how the
situation is different in 6D.

\subsection{Two-form fields in six dimensions}

We now come to the theory of two-form gauge fields in six dimensions.
This case is almost identical to the two-dimensional example of chiral
bosons we started with.  More generally, a similar situation arises in
a theory of self-dual and anti-self-dual fields in any dimension of
the form $4n+2$.  The conclusion in all these cases is that the charge
lattice $\Gamma$ must be integral and self-dual (but not necessarily
even).
Our analysis
relies only on the structure of the free field theory of two-form
fields, and is unchanged by interactions or coupling to gravity.  We
discuss the situation in the context of gravity in more detail in
Section \ref{sec:conclusions}.

As in two dimensions, we consider $p$ fields $B^{+a}$
whose field strength $H^{+a}=dB^{+a}$ is self-dual and $q$ fields $
B^{-r}$ whose field strength $H^{-r}=dB^{-r}$ is anti-self-dual.
The analog of the local operators (\ref{eq:twodop}) of the two-dimensional theory and the line operators (\ref{eq:Wilson}) of the
four-dimensional theory are the surface operators
\begin{equation}
W_{v_a, \tilde v_r}=\exp\left(i  \oint_{\Sigma_2}( v_a B^{+a} +  \tilde v_r B^{-r} ) \right)~.
    \label{eq:Wilsonsd}
\end{equation}
These operators describe the effect of a probe string with worldsheet
$\Sigma_2$ that carries charges $(v_a, \tilde v_r)$.  Because of the
(anti)self-dual properties of the fields, there is no need to add
independent 't Hooft operators.

Dirac quantization, or equivalently locality of the operators
(\ref{eq:Wilsonsd}), leads
to \cite{Deser-quantum}
\begin{equation}
v_a u_a - \tilde v_r \tilde u_r \in {\Z}~.
\label{eq:integrallas}
\end{equation}
This condition is very similar to (\ref{eq:integralla}).  It states that the signature $(p,q)$ lattice of charges $\Gamma$ has to be an integral lattice.

As in the two-dimensional case above, additional restrictions on the lattice $\Gamma$ cannot be seen by studying the theory on $\R^6$.  However, imposing consistency of the theory once it is compactified on various spaces shows that \begin{equation}
\Gamma= \Gamma^*~
\label{eq:self-duals}
\end{equation}
{\it i.e.} $\Gamma$ should be self-dual, as we now demonstrate.

Following the approach taken in \cite{Witten-dual}, we can relate the
6D theory to a 2D theory by compactifying the 6D theory on
${\C\P}^2$.  The resulting two-dimensional theory is precisely the
two-dimensional example we studied at the beginning of this section.
Each self-dual/anti-self-dual two-form field reduces to a
left/right-moving scalar field in 2D.  The operators \eq{eq:Wilsonsd}
reduce to operators \eq{eq:twodop}, with a corresponding reduction of
the charges $(v_a, \tilde{v}_r)$ to elements of the 2D lattice
$\Gamma$.  The self-duality of the charge lattice of the
two-dimensional theory leads to the conclusion (\ref{eq:self-duals})
for the self-duality of the lattice of the six-dimensional theory.

This simple but complete argument shows that a 6D field theory with
self-dual/anti-self-dual two-form fields must have a charge lattice
satisfying \eq{eq:self-duals}.  The condition that the charge lattice
of any six-dimensional theory is self-dual is, however, central to our
analysis.  Therefore it makes sense to view this conclusion from
different perspectives.  We now describe several other
compactifications which also lead to the condition \eq{eq:self-duals}.

As one alternative approach, we could
attempt to imitate the argument in two dimensions, which was based on
the partition function on $\S^1\times \S^1$, by considering the
partition function of the 6D theory on $\S^3\times \S^3$.  Then,
invariance under the exchange of the two $\S^3$ factors could also
lead to the self-duality condition (\ref{eq:self-duals}).  This
analysis must differ in some aspects from the 2D story, since there is
no analogue of the Hamiltonian picture.  We have not completed the
details of such an approach, but this could give another interesting
perspective on this problem.

Next, we consider the theory compactified on various tori.  (A related compactification of the theory of a single self-dual field on $\T^6$ was considered in \cite{Dolan-Nappi}.)  As with chiral scalar fields in 2D,
there is a question of boundary conditions when compactifying a theory
of 6D chiral and anti-chiral two-form fields on a torus.  Indeed, much of the formalism developed in \cite{Witten-5}-\cite{Mooretalks} is designed to address issues of this kind.  These works show that there is a $\Z_2$ structure needed for $B$ fields in six dimensions analogous to the spin structure needed
for fermions in 2D.  In the following subsection we discuss an odd charge
lattice that requires this generalization, analogous to the presence
of operators of half-integral spin in 2D.  Such odd lattices still
satisfy \eq{eq:self-duals}.  There does not seem to be any way to use
nontrivial boundary conditions to accommodate non-self-dual lattices,
though it would be nice to have a rigorous proof of this assertion.

Perhaps the most intuitive route to deriving (\ref{eq:self-duals})
follows from considering the theory on ${\R}^4 \times {\T}^2$.  Each
field $B^{+a}$ or ${B}^{-r}$ in six dimensions leads to a single
Maxwell field in four dimensions, and we can use these Maxwell fields
to describe the operators.  One set of operators is obtained by taking
the worldsheet $\Sigma_2 $ in (\ref{eq:Wilsonsd}) to be spanned by one
of the cycles in the compact ${\T}^2$ and a loop $\Sigma_1$ in
${\R}^4$.  This leads to a Wilson loop (\ref{eq:Wilson}) in four
dimensions.  If instead, we let $\Sigma_2$ wrap another cycle in
${\T}^2$, we find an 't Hooft loop in four dimensions.  Therefore, the
lattice $\Gamma$ of the six-dimensional theory includes both the
electric and the magnetic charges of the four dimensional theory.
Since in four dimensions the lattice of electric charges must be dual
to the lattice of magnetic charges (\ref{eq:Diracqua}), we see again
that the original six-dimensional lattice must be self-dual.  We give
a more explicit description of how the self-dual lattice reduces to
the electromagnetic charge lattice in 4D in terms of fields in the
following subsection.

We can also consider compactification of the 6D theory on
$\T^4$.  In this case we get several copies of the 6D charge lattice
in 2D from windings of each $H^\pm$ on different two-cycles in $\T^4$.
The resulting 2D charge lattice is a tensor product of the 6D charge
lattice with the unimodular lattice given by $H_2 (\T^4,\Z)$ with the
associated intersection form.  The condition
that the 2D charge lattice is unimodular again shows self-duality of
the 6D charge lattice.
Note that a similar argument for the 4D
theories considered above compactified on $\T^2$ gives an alternate
proof that all 4D charge lattices have an inner product of the form
(\ref{eq:4D-matrix}).

The arguments presented here show from
several points of view that any field theory with self-dual and/or
anti-self-dual two-form fields in six dimensions must have a self-dual
dyonic string charge lattice.  In particular, the dimensional
reductions on $\C\P^2$ and tori give simple yet complete
demonstrations of this result.

\subsection{Explicit example: $(1, 1)$ signature two-form fields}

It may be helpful to consider a specific example in more detail.
Consider the case of one $B^+$ and one $B^-$ in six dimensions, {\it
  i.e.} $(p, q) = (1, 1)$.  As discussed above, each of the two fields
$B^\pm$ reduces on $\T^2$ to a single $U(1)$ gauge field in four
dimensions.  We denote the compactified dimensions by $ 4$, $5$, and
note that the (anti)-self-duality condition states that the 6D field
strengths $H^\pm = dB^\pm$ satisfy
\begin{equation}
H_{4 \mu \nu}^{\pm} = \pm {1\over 2}\epsilon_{\mu \nu}^{\;\;\; \; \rho \sigma}
H_{5 \rho \sigma}^\pm \, .
\end{equation}
Thus, if we denote the field strength of the corresponding $U(1)$ in 4D by
\begin{equation}
F_{\mu \nu} = H_{4 \mu \nu}~,
\end{equation}
then strings wrapped around the 4-direction of the torus (which carry
string charge under $H_{4 0i}$, with $i$ in the radial spatial
direction) carry electric charge under the 4D $U(1)$, while strings
wrapped around the 5-direction carry magnetic charge under the same
$U(1)$.  Since strings wrapped in either direction carry charges in the
same 6D charge lattice $\Gamma$, the resulting theory in 4D has a single $U(1)$ field coming from the dimensional reduction of each of the two 6D $B$
fields.  The bilinear form on allowed electric and magnetic charges for
the resulting $U(1)^2 $ gauge group is then
\begin{equation}
\Gamma^{4D} =
\left(\begin{array}{cc}
0 & \Gamma\\
-\Gamma & 0
\end{array}
\right) \,.
\label{eq:gamma-reduced}
\end{equation}
As we noted above, the determinant of the skew-symmetric charge
product in 4D is always 1.  It follows that $\det \Gamma = \pm 1$, so
$\Gamma = \Gamma^*$.  By performing a linear transformation on
magnetic charges to put them in the dual basis to the electric
charges, it is always possible to transform \eq{eq:gamma-reduced} into
the form \eq{eq:4D-matrix}.

There are two distinct choices for the integral lattice $\Gamma$ with
signature $(1, 1)$.  We can have the odd lattice
\begin{equation}
\Gamma_1 = \left(\begin{array}{cc}
1 & 0\\
0 & -1\\
\end{array} \right).
\label{eq:gamma-odd}
\end{equation}
Or we can have the even lattice
\begin{equation}
\Gamma_0 = \left(\begin{array}{cc}
 0& 1\\
1 & 0\\
\end{array} \right).
\label{eq:gamma-even}
\end{equation}
Note that in the even case the two fields $B^\pm$ can be combined into
a single $B$ field with a Lagrangian description analogous to that of
a single 4D massless vector.  In this case, the analogue of Wilson and
't Hooft surface operators for the $B$ field are adequate
for describing the theory.  In the odd case, however, we must use the
surface operators for $B^\pm$ in eq.\ (\ref{eq:Wilsonsd}). Related to that is the fact that upon compactification the odd theory needs additional data like spin structure.

The 6D case should be contrasted with the situation in 4D.  In six dimensions, as in two dimensions, the dyonic charge lattice has a symmetric inner
product, and both \eq{eq:gamma-odd} and \eq{eq:gamma-even} are
possible charge lattices.  In four dimensions, the bilinear form on
the electric-magnetic charge space is antisymmetric, and there is only
one possible inner product structure on electric and magnetic charges.

Each of the two lattices $\Gamma_{1}, \Gamma_0$ is realized in certain
classes of string compactifications.  For $\Gamma_1$, after
dimensional reduction to 4D the bilinear form on charges is
\begin{equation}
 \Gamma^{{\rm 4D}}_1 =
\left(\begin{array}{cccc}
0 & 0 & 1 & 0\\
0 & 0 & 0 & -1\\
-1 & 0 & 0 & 0\\
0 & 1 & 0 & 0
\end{array} \right) \,.
 \label{eq:4D-odd}
\end{equation}
The analogous form arising from $\Gamma_0$ is
\begin{equation}
 \Gamma^{{\rm 4D}}_0 =
\left(\begin{array}{cccc}
0 & 0 & 0 & 1\\
0 & 0 & 1 & 0\\
0 & -1 & 0 & 0\\
-1 & 0 & 0 & 0
\end{array} \right) \,.
 \label{eq:4D-even}
\end{equation}
In each case, the matrix describing the form can be
put  into the canonical 4D form of
eq.\ (\ref{eq:4D-matrix}) by a change of basis for the magnetic charges.
For eq.\ (\ref{eq:4D-odd}) this can be done by changing the sign on
the fourth charge, while for eq.\ (\ref{eq:4D-even})
this is done by switching the third and fourth charges.

\section{Application to 6D supergravity}
\label{sec:SUGRA}

In six dimensions, antisymmetric two-index tensor fields appear in two distinct ${\cal N} = 1$ supersymmetry multiplets.  The gravity multiplet
contains a two-form field $B^+$ with a self-dual field strength.
Theories may also have any number $T$ of tensor multiplets each containing a two-form $B^-$ with an anti-self-dual field strength.  The set
of allowed charges carried by dyonic string excitations of the theory
must live in a lattice $\Gamma$ of signature $(1, T)$.  There is no
guarantee that all allowed dyonic charges on this lattice are realized
in the spectrum of the quantum theory, though strong arguments suggest that in a consistent theory of gravity there must be quantum
excitations realizing all allowed charges in the charge lattice (see e.g.\
\cite{Polchinski-completeness, Banks-Seiberg-completeness}).

In a general ${\cal N} = 1$ 6D supergravity theory, in addition to the
gravity multiplet and $T$ tensor multiplets there are $V$ vector
multiplets describing a general gauge group $G$ with simple
non-Abelian factors $G_1 \times \cdots \times G_k$, and $H$
hypermultiplets with scalars transforming in a representation of $G$.
The presence or absence of Abelian factors does not affect the
discussion here, which depends only on the non-Abelian part of the
gauge group.

We now review the Green-Schwarz anomaly cancellation conditions for 6D
supergravity theories \cite{Green-Schwarz-West, Sagnotti, Sadov, Honecker}, following \cite{tensors}.
The anomaly cancellation condition can be written in terms of the 8-form anomaly polynomial as
\begin{equation}
I_8(R,F) = \half \Omega_{\alpha\beta} X^\alpha_4 X^\beta_4.
\label{eq:factorized-anomaly}
\end{equation}
Here
\begin{equation}
X^\alpha_4 = \half a^\alpha \tr R^2 +  \sum_i b_i^\alpha \
\left(\frac{2}{\lambda_i} \tr F_i^2 \right)
\label{eq:string-current}
\end{equation}
with $a^\alpha, \ b_i^\alpha$ transforming as vectors in the space
$\R^{1,T}$ with symmetric inner product $\Omega_{\alpha\beta}$; ``$\tr$''
denotes the trace in the fundamental representation, and $\lambda_i$
are normalization constants depending on the type of each
simple group factor  ($\lambda = 1$ for $SU(N)$ factors).
Cancellation of the individual terms in \eq{eq:factorized-anomaly}
gives
\begin{eqnarray}
H-V & = &  273-29T\\
0 & = &     B^i_{\rm adj} - \sum_{\bf R}
x^i_{\bf R} B^i_{\bf R} \label{eq:f4-condition}\\
a \cdot a & =   &9 - T  \label{eq:aa-condition}\\
a \cdot b_i & =  & \frac{1}{6} \lambda_i  \left( A^i_{\rm adj} - \sum_{\bf R}
x^i_{\bf R} A^i_{\bf R}\right)  \label{eq:ab-condition}\\
b_i\cdot b_i & =  &-\frac{1}{3} \lambda_i^2 \left( C^i_{\rm adj} - \sum_{\bf R} x_{\bf R}^i C^i_{\bf R}  \right) \\
b_i \cdot b_j & = &  \lambda_i \lambda_j \sum_{\bf R S} x_{\bf R S}^{ij} A_{\bf R}^i
A_{\bf S}^j\label{eq:bij-condition}
\end{eqnarray}
where  $A_{\bf R},
B_{\bf R}, C_{\bf R}$ are group theory coefficients defined through
\begin{align}
\tr_{\bf R} F^2 & = A_{\bf R}  \tr F^2 \\
\tr_{\bf R} F^4 & = B_{\bf R} \tr F^4+C_{\bf R} (\tr F^2)^2 \,,
\end{align}
and where
$x_{\bf R}^i$ and $x_{\bf R S}^{ij}$
denote the number of matter fields that transform in the irreducible
representation $\bf R$ of gauge group factor $G_i$,
and $({\bf R} , {\bf S})$ of $G_i \otimes G_j$ respectively.
Note that for groups such as $SU(2)$ and $SU(3)$, which lack a fourth
order invariant, $B_{\bf R} = 0$ and there is no condition
\eq{eq:f4-condition}.

It is shown in \cite{tensors} using elementary group theory
that the inner products on the LHS of
conditions (\ref{eq:aa-condition}-\ref{eq:bij-condition}) are all
integral as a consequence of global and local anomaly cancellation.
Thus, independent of any UV realization of the theory, any apparently
consistent low-energy 6D supergravity theory contains an integral
lattice $\Lambda$ formed from vectors $a, b_i \in\R^{1, T}$.
We refer to this lattice as the {\em anomaly lattice} of the theory.
The
vector $a$ is associated with a coupling $B \cdot a \; \tr R^2$ of the
$B$ fields to space-time curvature, while the vectors $b_i$ are
associated with couplings $B \cdot b_i\;\tr F_i^2$ of the $B$ fields
to the field strengths $F_i$ of the various factors in the gauge group.

The 6D theory has gauge dyonic string excitations with charge $b_i$
associated with instantons in the factors of the gauge group
\cite{Duff-gauge}.  The charges $b_i$ therefore are contained within
the dyonic string charge lattice of the 6D theory.
While we do not have a complete argument that the charge $a$ also
corresponds to a dyonic string\footnote{We thank Ashoke Sen for useful
discussions on this point.}, there are several heuristic arguments
that suggest that this is the case.
Locally, a gravitational instanton will carry charge
$a$ and should live in the charge lattice.  The fact that $a$ has an
integer inner product with itself and all vectors $b_i$ indicates that
it satisfies Dirac quantization with respect to the other charges,
which also strongly suggests that it should also be an element of the
charge lattice.

F-theory \cite{Vafa-f, Morrison-Vafa-I, Morrison-Vafa-II} gives a very
general approach to constructing 6D ${\cal N} = 1$ supergravity
theories by describing nonperturbative IIB string compactifications on
a complex surface ${\cal B} $ containing 7-branes.  Geometrically,
monodromies around the 7-branes describe a Calabi-Yau threefold that
is elliptically fibered over the base ${\cal B} $.  For any F-theory
construction, the charge lattice $\Gamma =H_2 ({\cal B} ,\Z)$ is
spanned by dyonic strings in six dimensions arising when 3-branes are
wrapped on the 2-cycles of ${\cal B} $.  Self-duality of $\Gamma$ is
then an automatic consequence of Poincar\'{e} duality.
The lattice $\Lambda$ spanned by the vectors $a, b_i$ is embedded in a
simple fashion in the lattice $\Gamma$ for any F-theory model
\cite{tensors}.  In this embedding, $a$ corresponds to the canonical
class of the space ${\cal B} $ and is always associated with a state
in the charge lattice.

The condition that the lattice of charges $\Gamma$ for the two-form
fields of any 6D ${\cal N} = 1$ theory must be self-dual is a strong
constraint on 6D supergravity theories.  This constraint rules out a
large class of models that satisfy the other known consistency
constraints of anomaly cancellation and proper-sign gauge kinetic
terms.  In particular, this constraint implies that the anomaly
lattice $\Lambda$ should admit an embedding into a self-dual lattice
$\Gamma$.  In \cite{KMT, tensors} several theories were identified
that have anomaly lattices $\Lambda$ violating this
condition.  The result of the previous section shows that these
theories are inconsistent.  Thus, these theories are
not consistent quantum supergravity theories for any type of UV
completion.  In the following section we give some examples of such
theories.

\section{Examples of inconsistent theories}
\label{sec:examples}

We give several examples of theories that satisfy anomaly cancellation
and gauge kinetic term sign constraints, but that violate the
self-dual charge lattice condition.  In the first example we assume
that the charge $a$ is contained in the charge lattice.  The second
example is very similar but contains a charge lattice that is
disallowed even when $a$ is not considered.  The third example is a
model with two tensor multiplets, and the final example contains a
more exotic matter representation.  The examples given have $T = 1$
and $T = 2$.  For $T = 0$ the only integral lattice is $\Z$, which is
automatically self-dual, so no models in this class are ruled out by
the self-duality constraint.  For $T > 1$ it seems likely that a large
fraction of anomaly-free models violate the self-duality constraint,
though it is more difficult to prove in cases where the number of
simple factors in the gauge group is less than $T$.
\vspace*{0.05in}

\noindent {\bf Example 1:}

One example of a problematic model, first identified in \cite{KMT}
and discussed further in \cite{tensors}, is a 6D supergravity theory
with $T = 1$, gauge group
\begin{equation}
G = SU(4)
\end{equation}
and matter content
\begin{equation}
1 \times ({\bf 15})+
40 \times ({\bf 4})+
10 \times ({\bf 6})+
24 \times ({\bf 1})
\,.
\end{equation}
It is easy to check that $H-V = 244$ as required by anomaly
cancellation.  The anomaly lattice for
this theory consists of vectors $a, b$ spanning the lattice
\begin{equation}
\Lambda  = \left(\begin{array}{cc}
a \cdot a & -a \cdot b\\
-a \cdot b &  b \cdot b
\end{array} \right)
= \left(\begin{array}{cc}
8 & 10\\
10 & 10
\end{array} \right) \,.
\label{eq:sick-lambda}
\end{equation}
This lattice cannot be embedded in any unimodular lattice of signature
$(1, 1)$.
The easiest way to see this is to note that if $\Lambda$ is embedded
in a unimodular lattice of the same dimensionality as itself,
then the volume of the unit cell spanned by
$-a, b$ in $\Lambda$ must be an integer.  But the determinant of
$\Lambda$ is the volume squared, and therefore must be a perfect
square (up to a sign).  In this case $\det \Lambda = -20$, which is
not a perfect square, so $\Lambda$ does not admit an embedding into
any two-dimensional unimodular lattice.

We can also directly prove this assertion by examining all cases as
follows.  The only signature $(1, 1)$ unimodular lattices are
$\Gamma_{0,1}$ from (\ref{eq:gamma-even}, \ref{eq:gamma-odd}).  In the
case of $\Gamma_0$, the only choices for $a\cdot a = 8$ are (up to
overall sign) $(-2, -2)$ and (up to sign and exchange of coordinates)
$(-1, -4)$.  If $a=(-2, -2)$ then $b = (x, y)$ with $2xy = 10, 2 (x +
y) = 10$, which have no integral solutions for $x, y$.  Similarly, if
$a=(-1, -4)$ then $2xy = 10, x + 4y = 10$, which again has no integer
solutions.  For $\Gamma_1$, the only possibility (up to signs) for $a$
is $a =(-3, 1)$, which gives $x^2 -y^2 = 10, 3x + y = 10$, which once
again has no integer solutions.  Thus, $\Lambda$ cannot be realized by
any set of vectors in a unimodular lattice $\Gamma$, so this 6D theory
cannot be consistent when reduced on a torus to 4D, and therefore is
inconsistent in 6D.
\vspace*{0.05in}

\noindent {\bf Example 2:}

We now consider a closely related example, where the charge $a$ is not
needed to rule out the lattice.  Basically, the idea is to construct a
theory like the previous example, but with a second gauge group factor
having a dyonic string instanton with charge proportional to $a$.

In this example the gauge group is
\begin{equation}
G = SU(3) \times SU(3) \,.
\end{equation}
The matter content is
\begin{equation}
15 \times ({\bf 3},  {\bf 1}) +
45 \times ({\bf 1},  {\bf 3})+
5 \times ({\bf 3},  {\bf 3})+
1 \times ({\bf 1},  {\bf 8})+
27 \times ({\bf 1},  {\bf 1})\,.
\end{equation}
It can be verified that both
local and global anomalies cancel for this model, and the anomaly
lattice is spanned by vectors $-a, b_1, b_2$ with inner products
\begin{equation}
\Lambda= \left(\begin{array}{ccc}
8 & 4 &10\\
4 & 2 & 5\\
10 & 5 &10
\end{array} \right) \,.
\label{eq:sick-lambda-2}
\end{equation}
Note that this lattice is degenerate, with $b_1 = -a/2$ and hence we can focus on the two-dimensional lattice spanned by the dyonic string charges $b_1, b_2$,
\begin{equation}
\hat{\Lambda}= \left(\begin{array}{ccc}
 2 & 5\\
 5 &10
\end{array} \right) \,.
\label{eq:sick-lambda-21}
\end{equation}
The arguments used in the previous example show that this lattice cannot
be embedded in any unimodular lattice, so the theory is inconsistent.
\vspace*{0.05in}

\noindent {\bf Example 3:}

For another example, take the theory with $T = 2$
with gauge group
\begin{equation}
G = SU(N) \times SU(N) \,.
\end{equation}
and charged matter content
\begin{equation}
2 N \times ({\bf N},  {\bf 1}) +
2 N \times ({\bf 1},  {\bf N})
\,.
\end{equation}
The anomaly
lattice for this model is spanned by vectors $-a, b_1, b_2$ with inner products
\begin{equation}
\Lambda= \left(\begin{array}{ccc}
7 & 0 &0\\
0 & -2 & 0\\
0 &  0 & -2
\end{array} \right) \,.
\label{eq:sick-lambda-3}
\end{equation}
Once again, this does not admit a unimodular embedding since the
determinant is 28, which is not a perfect square.  This is another
example in which the inconsistency of the model depends upon the
appearance of $a$ in the charge lattice.  (Note for comparison that
the theory with one $SU(N)$ factor and $2 N$ fundamental representations
with $T = 1$ has an anomaly lattice diag$(8, -2)$ with determinant -16,
which does admit an embedding into a unimodular lattice.)

\noindent {\bf Example 4:}

For another example, take the theory with $T = 1$
with gauge group
\begin{equation}
G = SU(3) \times SU(4) \,.
\end{equation}
and matter content
\begin{equation}
2 \times ({\bf 3},  {\bf 4}) +
22 \times ({\bf 3},  {\bf 1})+
13 \times ({\bf 1},  {\bf 4})+
1 \times ({\bf 1},  {\bf 20'} \left[ \tiny\yng(2,1) \right])
+105 \times ({\bf 1},  {\bf 1})
\,.
\end{equation}
The anomaly
lattice for this model is spanned by vectors $-a, b_1, b_2$ with inner products
\begin{equation}
\Lambda= \left(\begin{array}{ccc}
8 & 4 &4\\
4 & 2 & 2\\
4 & 2 &6
\end{array} \right) \,.
\label{eq:sick-lambda-4}
\end{equation}
As in example 2, this lattice is degenerate, with $b_1 =
-a/2$.  Considering only the lattice spanned by dyonic string charges
$b_1, b_2$, a similar argument to the other examples shows
that the lattice cannot be embedded in any unimodular lattice, so the
theory is not consistent.

\section{Concluding remarks}
\label{sec:conclusions}

Following on the work of \cite{Witten-5}-\cite{Mooretalks} we have
given simple physical arguments showing that the charge lattice of
two-form fields in six dimensions should be integral and self-dual
(but not necessarily even).  The criteria we have used, based on
consistency of the theory upon compactification to lower dimensions,
can be interpreted as physical requirements for the theory.  Despite
the substantial progress of \cite{Witten-5}-\cite{Mooretalks}, we feel
that a simpler mathematical treatment connecting the general formalism
to the physical properties of these theories is still desirable.  But
even without such a deeper understanding, we can use these consistency
conditions to derive powerful constraints on supergravity theories and
therefore also on possible string constructions.  The results of this
paper show that a certain class of 6D supergravity theories are
inconsistent.  This narrows the gap between the set of 6D theories
that are not provably inconsistent and the set of theories with known
string realizations.

The arguments given in Section \ref{sec:consistency} for self-duality
of the 6D charge lattice were based  on principles of field
theory.  It is important to stress that while coupling to gravity can
lead to additional consistency conditions, the conditions derived from
field theory are still necessary in the presence of gravity and other
fields.  In particular, when we consider compactification of a theory
with two-forms and gravity on $\C\P^2$ we can take the size of the
compact space to be much larger than the Planck length, so that all
gravitational effects are negligible.

When compactifying on tori there is also the
issue that new charges appear coupling to the gauge fields arising
from the 6D metric.  This extends the charge lattice of the lower
dimensional theory.  Kaluza-Klein momenta produce electric charges for
these additional gauge fields.
We can choose, however, to perform the reduction on a square torus (or
$d$-dimensional generalization thereof) in
the absence of background fields, in which case
the resulting extension of the charge lattice is only by a direct sum
with a trivial $2d \times 2d$ matrix of the form \eq{eq:4D-matrix}.
This does not change the determinant, so the unimodularity constraint
still holds when gravity is added.

KK monopoles can also appear which naively will carry a dyonic charge
associated with the vector $a$ appearing in the coupling $a \cdot B
R^2$.  This does not change the self-dual structure of the charge
lattice, but does give another reason to suspect that $a$ is in the
charge lattice.  The role of the dyonic charge $a$ in the theory is,
however, somewhat subtle.  Although this charge was shown in
\cite{tensors} to satisfy Dirac quantization with respect to gauge
dyonic strings for any theory satisfying anomaly cancellation, we do
not have a rigorous proof that this dyonic charge must appear in the
charge lattice.  It would be nice to have a more complete
understanding of the role of the dyonic string charge $a$.  One promising
approach to understanding this and related questions is through a
systematic understanding of the world-sheet theory on the dyonic
strings themselves.

We have used the condition that the dyon lattice must be self-dual to rule out a number of 6D supergravity theories which otherwise obey the known consistency conditions.  In some cases, the gauge instantons in the 6D theory provide
a set of dyonic string charges that is sufficient to rule out the theory.
In other cases, the full anomaly lattice cannot be embedded
in a self-dual lattice only because of the vector $a$ associated with
gravitational physics.  In the latter set of cases, a better
understanding of $a$ is needed to rigorously rule out these theories.

One might also wonder whether the modifications of the sum over
topological sectors of \cite{Seiberg:2010qd} and
\cite{Banks-Seiberg-completeness} can affect our conclusions.  These
papers coupled a discrete gauge theory to the topological charges,
which has the effect of excluding certain topological sectors.  We can
contemplate coupling the various four-form string currents in
\eq{eq:string-current} to such a discrete gauge theory.  This amounts
to enlarging the gauge symmetry of the two-forms $B^\pm$ by adding a
discrete gauge symmetry for such forms.  One way to think about such discrete gauge symmetries \cite{Maldacena:2001ss, Banks-Seiberg-completeness} is to embed them in continuous symmetries for additional $B$ fields and then Higgs them down to the discrete symmetry.  For example, if the discrete gauge symmetry is $\Z_p$, we add to the theory a two-form field $B$, which includes both a self-dual and an anti-self-dual $H=dB$, as well as its Higgs field, which is a one-form $A$.  The presence of the new $B$ fields enlarges the charge lattice.  But since the added fields are ``vector like'', they cannot affect our conclusion about the original system and
the charge lattice of the massless $B^\pm$ should still be self-dual.

\vspace*{0.2in}

\noindent
{\bf Acknowledgements}: We thank Ashoke Sen for collaboration in the
early stages of this project. We would also like to thank Tom Banks,
Frederik Denef, Vijay Kumar, Greg Moore, David Morrison, Daniel Park
and Edward Witten for helpful discussions.  We are particularly
thankful to Greg Moore for his patient explanations of his work on
Abelian gauge fields.  This research was supported in part by the DOE
under contracts \#DE-FC02-94ER40818 and DE-FG02-90ER40542.

\end{document}